\documentstyle[epsfig,12pt,equation]{article}
\begin{document}
\def\to{\rightarrow}
\def\ub{\underbar}
\def\rs{\mbox{$\sqrt{s}$}}
\def\gga{\mbox{$G^{\gamma}$}}
\def\jpsi{\mbox{$J/\psi$}}
\def\lqcd{\mbox{$\Lambda_{\rm QCD}$}}
\def\ppbar{\mbox{$p \bar{p}$}}
\def\qqbar{\mbox{$q \bar{q}$}}
\def\QQbar{\mbox{$Q \bar{Q}$}}
\def\bbbar{\mbox{$b \bar{b}$}}
\def\ccbar{\mbox{$c \bar{c}$}}
\def\mcc{\mbox{$M_{c \bar{c}}$}}
\def\ttbar{\mbox{$t \bar{t}$}}
\def\ben{\begin{subequations}}
\def\be{\begin{equation}}
\def\een{\end{subequations}}
\def\ee{\end{equation}}
\def\beq{\begin{eqalignno}}
\def\eeq{\end{eqalignno}}
\def\bea{\begin{eqnarray}}
\def\eea{\end{eqnarray}}
\def\qsq{\mbox{$Q^2$}}
\def\xgam{\mbox{$x_{\gamma}$}}
\def\sigtot{${\sigma^{tot}_W}$}
\def\gamg{${ \gamma g }$}
\def\ptmin{\mbox{$ p_{T,{\rm min}} $}}
\def\egam{\mbox{$e \gamma $}}
\def\gaga{\mbox{$\gamma \gamma $}}
\def\epem{\mbox{$e^+ e^- $}}
\def\alphas{\mbox{$\alpha_s$}}
\def\aem{\mbox{$\alpha_{\rm em}$}}
\def\rts{\mbox{$ \sqrt{s} $ }}
\def\gamp{\mbox{$\gamma p$}}
\def\totwo{$2 \rightarrow 2$}
\def\glph{\mbox{$ G^{\gamma}(x,Q^2)$}}
\def\Gg{\mbox{$ G^{\gamma}$}}
\def\ug{\mbox{$ u^{\gamma}$}}
\def\dg{\mbox{$ d^{\gamma}$}}
\def\sg{\mbox{$ s^{\gamma}$}}
\def\cg{\mbox{$ c^{\gamma}$}}
\def\qvph{\mbox{$\vec q^{\gamma} $}}
\def\qv0{\mbox{$\vec q^{\gamma}_0 $}}
\def\qph{\mbox{${q^{\gamma} } $}}
\def\vqxqsq{\mbox{$ \vec q^{\gamma} (x,Q^2)  $}}
\def\qisq{\mbox{$ q_i^{\gamma} (x,Q^2) $}}
\def\qolsq{ (Q^2 / \Lambda ^2)}
\def\fgme{\mbox{$f_{\gamma|e}$}}
\def\fgmebeam{\mbox{$ f_{\gamma|e}^{beam}$ }}
\def\fgmer{\mbox{$ f_{\gamma|e}^{res} $}}
\def\sigh{ \hat \sigma}
\def\xtsq{\mbox{$ x{_T}{^2}$}}
\def\f2gam{\mbox{$ F_2^\gamma $}}
\def\flgam{\mbox{$ F_L^\gamma $}}
\renewcommand{\thefootnote}{\fnsymbol{footnote}}
\hyphenation{brems-strahl-ung beam-strahl-ung}
\setcounter{footnote}{0}
\begin{flushright}
CTS-TH-1/96\\
February 1996\\
hep-ph/9602428\\
\end{flushright}
\vspace{1cm}
\begin{center} 
{\Large \bf Photon Structure Function
\footnote{Invited talk at the XI DAE Symposium, Santiniketan, Jan. 95}
}\\
\vspace{1cm}
Rohini M. Godbole\footnote{On leave from Dept. of Physics, University of 
Bombay, Bombay,India.}\footnote{E-mail : rohini@cts.iisc.ernet.in} \\
{\it Center for Theoretical Studies, Indian Institute of Science, 
Bangalore, 560 012, India.}\\
\vspace{1cm}
\end{center}
\begin{abstract}
After  briefly  explaining the idea of photon structure functions (\f2gam\ ,
\flgam) I review the  current theoretical and experimental developements in 
the subject of extraction of \qvph\ from a study of the Deep
Inelastic Scattering (DIS).  I then end by pointing out 
recent progress in getting  information about the parton content of the 
photon from  hard processes other than DIS.
\end{abstract}

\vspace{.5cm}

\noindent {\bf \large  Introduction:}

\vspace{.5cm}

The photon is the simplest of all bosons.
Quantum Electrodynamics (QED), the theory of $e-\gamma$ interactions 
is the most accurately tested field theory we have. 
At first sight therefore it is  surprising that many
reactions involving (quasi--)real photons are much less well understood, 
both theoretically and experimentally.  This  outwardly strange fact
is the result of fluctuations of a photon into quark--antiquark 
pairs. Whenever the lifetime of the virtual state exceeds the
typical hadronic time scale the (virtual) \qqbar\ pair has 
sufficient time to evolve into a complicated hadronic state that 
cannot be described by perturbative methods only. Even 
if the lifetime is shorter, hard gluon emission and related processes 
complicate the picture substantially. This thus endows the photon with 
a hadronic structure so to say.

The understanding of these virtual hadronic states becomes particularly 
important when they are ``kicked on the mass shell" by an interaction of the
photon. The most thoroughly studied reactions of this type involve 
interactions of a real and a virtual photon (e.g. in $e \gamma$ scattering);
of two real photons (\gaga\ scattering at \epem\ colliders), the so called 
Deep
Inelastic Scattering (DIS) off a photon target. There are two very important
reasons for us to be interested in the study of hadronic structure of the 
photon; one is to facilitate better understanding of the interactions of high
energy photons which can help sharpen our assesment of  backgrounds at
high energy colliders; this is true for high--energy linear \epem\ colliders 
that are now being discussed, and especially for the so--called \gaga\ 
colliders and second is the unique opportunity that the photon provides to 
study the perturbative and nonperturbative aspects of QCD. The latter is  
due to the fact that ``in principle'' the hadronic structure of the photon 
arises from the ``hard'' $\gamma q \bar q$ vertex.

I discuss below mainly the DIS. These were the first photonic reactions for 
which predictions were made in the framework of the quark parton model 
(QPM) \cite{2} and within QCD \cite{3}.  $e \gamma$ scattering was also among 
the first of the ``hard'' photonic reactions, which can at least partly 
be described by perturbation theory, to be studied experimentally \cite{4}. 
After that in the end I will discuss in brief how one 
can use the  `resolved photon' processes \cite{5} to probe the structure of 
the photon and  mention some new experimental data which fortells 
the expected progress in the area.

\vspace{.5cm}

\noindent {\bf \large Photon Structure Functions:}

\vspace{.5cm}

Deep--inelastic \egam\ scattering (DIS), is theoretically very clean, 
being fully inclusive; it is thus well suited to serve as the defining process
for photon structure functions and the parton content of the photon. 
See Ref.\cite{9a} for a pedagogical introduction to the subject.

Formally deep--inelastic \egam\ scattering is quite similar to $ep$ 
scattering 
\be \label{e2.1}
e \gamma \to e X,
\ee
where $X$ is any hadronic system and the squared four momentum transfer $Q^2
\equiv - q^2 \geq 1$ GeV$^2$.  The basic kinematics is explained in Fig.~1. 
\begin{figure}[htb]
\leavevmode 
\begin{center}
\mbox{\epsfxsize=0.25\hsize\epsffile{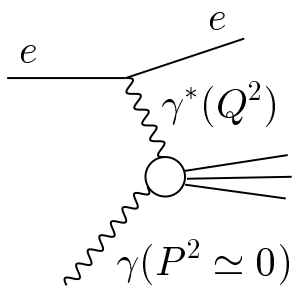}}
\caption{Deep Inelastic Scattering off a photon target.}
\end{center}
\end{figure}
The differential cross section can be written in terms of the scaling 
variables $x \equiv Q^2 / (2 p \cdot q)$ and $y \equiv Q^2 / (s x)$, 
where \rs\ is the total available centre--of--mass (cms) energy:
\be \label{e2.2}
\frac {d^2 \sigma (e \gamma \to e X)} {dx dy} = 
\frac {2 \pi \alpha^2_{\rm em} s} {Q^4}
\left\{ \left[ 1 + \left( 1-y \right)^2 \right] \f2gam(x,Q^2) - y^2 
F_L^{\gamma}(x,Q^2) \right\};
\ee
this expression is completely analogous to the equation defining the protonic
structure functions $F_2$ and $F_L$ in terms of the differential 
cross--section for $ep$ scattering via the exchange of a virtual photon.
The special significance \cite{2} of \egam\ scattering lies in the fact that,
while (at present) the $x-$dependence of the nucleonic structure functions can
only be parametrized from data, the structure functions appearing in
eq.(\ref{e2.2}) can be {\em computed} in the QPM from the diagram shown in 
Fig.~2a: 
\ben \label{e2.3} \beq
F^{\gamma,{\rm QPM}}_2(x,Q^2) &= \frac {3 \aem}{\pi} x \sum_q e_q^4 \left\{
\left[ x^2 + \left( 1-x \right)^2 \right] \log \frac {W^2} {m_q^2} + 8 x 
\left( 1-x \right) - 1 \right\},\label{e2.3a}\\
F^{\gamma,{\rm QPM}}_L(x,Q^2) &= \frac{3 \aem}{4 \pi} 
\sum_q e_q^4 4 x^2 (1-x).\label{e2.3b}
\eeq
\een
where we have introduced the squared cms energy of the $\gamma^* \gamma$ 
system
\be \label{e2.4}
W^2 = Q^2 \left( \frac {1}{x} - 1 \right).
\ee
The sum in eq.(\ref{e2.3}) runs over all quark flavours, and $e_q$ is the
electric charge of quark $q$ in units of the proton charge. Note that unlike
the case of the proton, for the photon $F_L^\gamma$ is nonzero even in the
QPM. 

\begin{figure}
\leavevmode
\begin{center}
\mbox{\epsfxsize=0.6\hsize\epsffile{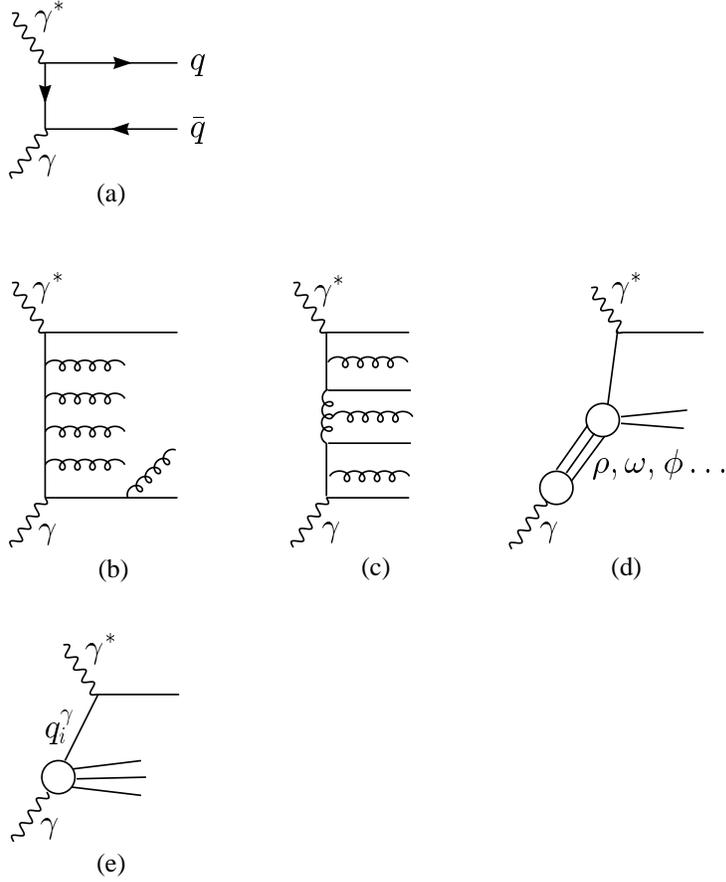}}
\end{center}
\caption{Different contributions to \protect\f2gam\ }
\end{figure}
Unfortunately, eq.(\ref{e2.3}) depends on the quark masses $m_q$. If this
ansatz is to describe data \cite{14} even approximately, one has to use
constituent quark masses of a few hundred MeV here; constituent quarks are not
very well defined in field theory. Moreover, we now know that QPM predictions
can be modified substantially by QCD effects. 
In case of \egam\ scattering, QCD corrections are described by the kind
of diagrams shown in Figs.~2b,c. Diagrams of the type 2b leave the flavour
structure unchanged and are therefore part of the (flavour) nonsinglet
contribution to \f2gam, while diagrams with several disconnected quark lines,
as in Fig.~2c, contribute to the (flavour) singlet part of \f2gam.

The interest in photon structure functions received a boost in 1977, when
Witten showed \cite{3} that such diagrams can be computed exactly, at least
in the so--called ``asymptotic" limit of infinite $Q^2$. Including 
next--to--leading order (NLO) corrections \cite{15}, the result can be
written as
\be \label{e2.5}
F_2^{\gamma,{\rm asymp}}(x,Q^2) = \aem \left[ \frac {1} {\alpha_s(Q^2)}
a(x) + b(x) \right],
\ee
where $a$ and $b$ are {\em calculable} functions of $x$. The absolute
normalization of this ``aymptotic" solution is therefore given uniquely by
$\alpha_s(Q^2)$, i.e. by the value of the QCD scale parameter $\Lambda_{\rm
QCD}$. It was therefore hoped that eq.(\ref{e2.5}) might be exploited for a
very precise measurement of $\Lambda_{\rm QCD}$. 

Unfortunately this no longer appears feasible. One problem is that, in order
to derive eq.(\ref{e2.5}), one has to neglect terms of the form $\left( \frac
{\alpha_s(Q^2)} {\alpha_s(Q_0^2)} \right)^P$, where $Q_0^2$ is some input
scale (see below). Neglecting such terms is formally justified {\em if}
$\alpha_s(Q^2) \ll \alpha_s(Q_0^2)$ {\em} and $P$ is positive. Unfortunately
the first inequality is usually not satisfied at experimentally accessible
values of $Q^2$, assuming $Q_0^2$ is chosen in the region of applicability of
perturbative QCD, i.e. $\alpha_s(Q_0^2)/\pi \ll 1$. Worse yet, $P$ can be zero
or even negative! In this case ignoring such terms is obviously a bad
approximation. Indeed, one finds that eq.(\ref{e2.5}) contains divergencies as
$x \to 0$ \cite{3,15}: 
\be \label{e2.6}
a(x) \sim x^{-0.59}, \ \ \ \ \ b(x) \sim x^{-1}.
\ee
The coefficient of the $1/x$ pole in $b$ is {\em negative}; eq.(\ref{e2.5})
therefore predicts negative counting rates at small $x$. Notice that the
divergence is worse in the NLO contribution $b$ than in the LO term $a$. It
can be shown \cite{16} that this trend continues in yet higher orders.
Clearly the ``asymptotic" solution is not a very useful concept, having a
violently divergent perturbative expansion.

The worst divergencies in $F_2^{\gamma,{\rm asymp}}$ occur in the singlet
sector, i.e. originate from diagrams of the type shown in Fig.~2c. 
There exist also non--perturbative contributions to $F_2^\gamma$ which 
are traditionally estimated using the vector dominance model (VDM)
\cite{18}, from the diagrams shown in Fig.~2d and we have
$$ 
F_2^{\gamma, {\rm VDM}} \propto F_2^{\rho, \omega, \phi} \simeq F_2^{\pi(p)}
$$
Hence one expects the contribution of Fig.~2d to be well--behaved, i.e.
non--singular. Hence this  {\em cannot} cancel the divergencies of the 
``asymptotic" solution.

This discussion tells us that we cannot hope to compute $\f2gam(x,Q^2)$ 
from perturbation theory alone.  
The only meaningful approach seems to be  that suggested by
Gl\"uck and Reya \cite{19}. That is, one {\em formally} sums the contributions
from Figs.~2a--d into the single diagram of Fig.~2e, where we have introduced
quark densities in the photon \qisq\ such that (in LO) 
\be \label{e2.7}
\f2gam(x,Q^2) = 2 x \sum_i e^2_{q_i} \qisq,
\ee
where the sum runs over flavours, $e_{q_i}$ is the electric charge of quark
$q_i$ in units of the proton charge, and the factor of 2 takes care of
anti--quarks. This is merely a definition. In the approach of ref.\cite{19}
one does not attempt to compute the absolute size of the quark densities 
inside the photon. Rather, one introduces input distribution functions
$q^{\gamma}_{i,0}(x) \equiv q_i^{\gamma}(x,Q_0^2)$ at some scale $Q_0^2$. 
$Q_0^2$ is usually chosen as the {\em smallest} value for which 
$\alpha_s(Q_0^2)$ is sufficiently small to allow for a meaningful 
perturbative expansion.  

Given these input distributions, the photonic parton densities, and thus 
\f2gam, at different values of $Q^2$ can be computed using the inhomogeneous
evolution equations. In LO, they read \cite{3,20}:
\ben \label{e2.8} \beq
\frac { d q^{\gamma}_{\rm NS} (x,Q^2)} { d \log Q^2} &= \frac {\aem}{2 \pi}
k^{\gamma}_{\rm NS}(x) + \frac {\alpha_s(Q^2)} {2 \pi} \left( P^0_{qq} \otimes
q^{\gamma}_{\rm NS} \right) (x,Q^2) ; \label{e2.8a} \\
\frac { d \Sigma^{\gamma} (x,Q^2)} { d \log Q^2} &= \frac {\aem}{2 \pi}
k^{\gamma}_{\Sigma}(x) + \frac {\alpha_s(Q^2)} {2 \pi} \left[ \left( P^0_{qq}
\otimes \Sigma^{\gamma} \right) (x,Q^2) + \left( P^0_{qG} \otimes G^{\gamma}
\right) (x,Q^2) \right] ; \label{e2.8b} \\
\frac { d G^{\gamma} (x,Q^2)} { d \log Q^2} &= \frac {\alpha_s(Q^2)} {2 \pi} 
\left[ \left( P^0_{Gq}
\otimes \Sigma^{\gamma} \right) (x,Q^2) + \left( P^0_{GG} \otimes G^{\gamma}
\right) (x,Q^2) \right], \label{e2.8c}
\eeq \een
where we have used the notation 
\be \label{e2.9}
\left( P \otimes q \right) (x,Q^2) \equiv \int_x^1 \frac {dy}{y} P(y)
q(\frac{x}{y},Q^2).
\ee
The $P^0_{ij}$ are the usual (LO) $j \to i$ splitting functions 
and $k_i^{\gamma}$ describe $\gamma \to q \bar q$ splitting.
Eq.(\ref{e2.8a}) describes the evolution of the nonsinglet distributions 
(differences of quark
densities), i.e. re--sums only diagrams of the type shown in Fig.~2b, while
eqs.(\ref{e2.8b},\ref{e2.8c}) describe the evolution of the singlet sector
($\Sigma^{\gamma} \equiv \sum_i q_i^{\gamma} + \bar{q}_i^{\gamma}$), which
includes diagrams of the kind shown in Fig.~2c. Notice that this necessitates
the introduction of a gluon density inside the photon \glph, with its
corresponding input distribution $G_0^{\gamma}(x) \equiv G^{\gamma}(x,Q_0^2)$.

It is crucial to note that, given non--singular input distributions, the
solutions of eqs.(\ref{e2.8}) will also remain \cite{19} well--behaved at all
finite values of $Q^2$. This is true both in LO and in NLO \cite{22}. On the
other hand one clearly has abandoned the hope to make an absolute prediction 
of $\f2gam(x,Q^2)$ in terms of $\Lambda_{\rm QCD}$ alone. 
The solutions of eqs.(\ref{e2.8}) still show an
approximately linear growth with $\log Q^2$; in this sense eq.(\ref{e2.5})
remains approximately correct, but the functions $a$ and $b$ now do depend
weakly on $Q^2$ (approximately like $\log \log Q^2$), and the $x-$dependence
of $b$ is {\em not} computable.  This approximate linear growth of \f2gam\ 
with \qsq\ has been experimentally confirmed quite nicely\cite{amy} as shown
in Fig.~3 taken from Ref. \cite{amy}.
\begin{figure}[htb]
\leavevmode
\begin{center}
\mbox{\epsfxsize=.5\hsize\epsffile{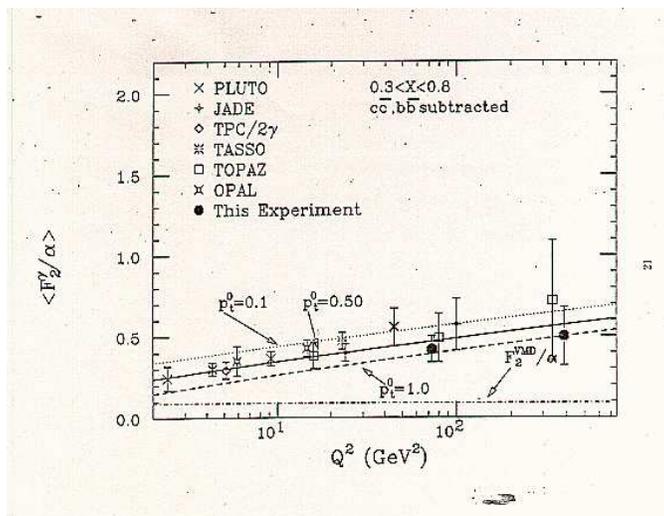}}
\end{center}
\caption{$x$ averaged data on \f2gam\ as a function of $Q^2$\protect\cite{amy}.}
\end{figure}
Notice that {\em no} momentum sum rule applies for the parton
densities in the photon as defined here. The reason is that these densities
are all of first order in the fine structure constant \aem. Even a relatively
large change in these densities can therefore always be compensated by a small
change of the ${\cal O}(\alpha^0_{\rm em})$ term in the decomposition of the
physical photon, which is simply the ``bare" photon [with distribution
function $\delta(1-x)$]. 

Before discussing our present knowledge of and parametrizations for the
parton densities in the photon, we briefly address a few issues related to the
calculation of \f2gam. As mentioned above, eqs.(\ref{e2.7}),(\ref{e2.8})
have been extended to NLO quite early, although a mistake in the two--loop
$\gamma \to G$ splitting function was found \cite{25} only fairly recently. A
full NLO treatment of massive quarks is now also available \cite{26} for both
\f2gam\ and $F_L^{\gamma}$. A first treatment of small$-x$ effects in the
photon structure functions, i.e. $\log 1/x$ re--summation and parton
recombination, has been presented in ref.\cite{27}; however, the predicted
steep increase of \f2gam\ at small $x$ has not been observed experimentally
\cite{28}. Finally, non--perturbative contributions to \f2gam\ are expected to
be greatly suppressed if the target photon is also far off--shell. One can
therefore derive unambiguous QCD predictions \cite{29} in the region $Q^2 \gg
P^2 \gg \Lambda^2$, where the first strong inequality has been imposed to
allow for a meaningful definition of structure functions.
However, it has recently been pointed out
\cite{30} that non--perturbative effects might survive longer than previously
expected; an unambiguous prediction would then only be possible for very large
$P^2$, and even larger $Q^2$, where the cross--section is very small.

\vspace{.5cm}

\noindent {\bf \large  Parametrizations of Photonic Parton Densities:}

\vspace{.5cm}

As discussed above  the $Q^2$ evolution of the photonic
parton densities $\qvph(x,Q^2) \equiv (q_i^{\gamma}$ , $G^{\gamma})(x,Q^2)$ 
is uniquely determined by perturbative QCD, eqs.(\ref{e2.8}) and their NLO
extension once input distributions \qv0\ at a fixed $Q^2 = Q_0^2$ are
specified. Though similar to the nucleonic case, the determination of the 
input distributions is much more difficult in case of the photon, for a 
variety of reasons. 

To begin with, no momentum sum rule applies for \qv0, as discussed above. This
means that it will be difficult to derive reliable information on 
$G_0^{\gamma}$ from measurements of $F_2^{\gamma}$ alone: in LO, the gluon
density only enters via the (subleading) $Q^2$ evolution in $F_2^{\gamma}$. 
We will see below that parametrizations for $G^{\gamma}$ still differ 
by sizable factors over the entire $x$ range unless $Q^2$ is very large.

Secondly, so far deep--inelastic $e \gamma$ scattering could only be studied
at \epem\ colliders, where the target photon is itself radiated off one of the
incoming leptons. The cross section from the measurement of which
$F_2^{\gamma}$ is to be determined is of order $d \sigma / d Q^2 \sim
\alpha^4_{\rm em}/ \left( \pi Q^4 \right) \log \left( E/m_e \right)$, see 
eq.(\ref{e2.2}). The event rate is therefore quite small; the most recent 
measurements \cite{14,28,32,amy} typically have around 1,000 events at 
$Q^2 \simeq 5$ GeV$^2$, and the statistics rapidly gets worse at higher 
$Q^2$. 

Another problem is that the $e^{\pm}$ emitting the target photon is usually
not detected, since it emerges at too small an angle. This means that the
energy of the target photon, and hence the Bjorken variable $x$, 
can {\em only}
be determined from the hadronic system. All existing analyses try to determine
$x$ from the invariant mass $W$, using eq.(\ref{e2.4}). 
Since at least some of the produced hadrons usually also escape undetected,
the measured value of $W$ ($W_{\rm vis}$) is generally {\em smaller}
than the true $W$. One has to correct for this by ``unfolding " the  
measured $W_{\rm vis}$) distribution to arrive at the true $W$ ( and 
hence true $x$ ) distributions. To do this one has to model the hadronic 
system $X$. Reasonably well tested alogrithms have been  evolved for this.
The procedure normally used by the experimentalists is as originally 
suggested in Ref. \cite{35}. However the way it is implemented currently
\cite{36,28,32,amy} has certain shortcomings \cite{5,36a}.
Also this  procedure can lead to large uncertainties at the boundaries of the
accessible range of $x$ values. It has been shown explicitly \cite{34} that 
different ans\"atze for $X$ can lead to quite different ``measurements" of 
$F_2^{\gamma}$ at small $x$. The estimation of systematic error due to 
unfolding procedure only includes things like the choice of binning \cite{28}
and not some of the uncertainites in the modelling of the state $X$.
This might help to explain the apparent discrepancy between different data 
sets \cite{14}. Fortunately, new ideas for improved unfolding algorithms 
\cite{36b} are now under investigation; this should facilitate the 
measurement of $F_2^{\gamma}$ at small $x$, especially at high energy (LEP2)
\cite{lep2gg}
In spite of this, measurements of $F_2^{\gamma}$ probably still provide the
most reliable constraints on the input distributions $\qv0(x)$; they are
certainly the only data that have been taken into account when constructing
existing parametrizations of \vqxqsq.

At present there exist a large number $( \sim 20)$ 
parametrisations for the photonic parton densities. Apart from the
simplest and the oldest parametrisations \cite{37,37a} based on 
``Asymptotic"  LO prediction \cite{3,37b}, (which were recently 
improved by Gordon and Storrow \cite{38}) all other parametrizations involve
some amount of data fitting. However, due to the rather large experimental 
errors of data on $F_2^{\gamma}$, additional {\em assumptions} always had to 
be made.  The different parametrisations are not just different fits
to the data but they differ from each other in these assumptions, the 
treatment of heavy quarks, choice of the scale $Q_0^2$ and the physics 
ideas used for this choice of input densities.
One assumption made by all of them is that quark and anti--quark
distributions (of the same flavour) are  identical, which
guarantees that the photon carries no flavour. 

The \ub{DG parametrization} \cite{39} was the first to start from input
distributions and is based on only a single measurement of \f2gam\ at 
$Q^2 \simeq 5.2 {\rm GeV}^2$ that was then available. Two assumptions 
were made : All input quark
densities were assumed to be proportional to the squared quark charges, i.e.
$\ug = 4 \dg = 4 \sg$ at $Q_0^2 = 1$ GeV$^2$; and the gluon input was
generated purely radiatively. This parametrization only exists in LO. 
The charm content is definitely overestimated in this parametrisation. 

The \ub{LAC parametrizations} \cite{40} are based on a much larger data set.
The main point of these fits  was to demonstrate that data on $F_2^{\gamma}$
constrain \Gg\ very poorly. In particular, they allow a very hard gluon,
(LAC3), as well as very soft gluon distributiuons (LAC1, LAC2).
The LAC parametrizations only exist for $N_f=4$ massless flavours and in LO.
{\em No} assumptions about the relative sizes of the four input quark
densities were made in the fit. LAC3 has been clearly excluded by data on jet
production in $ep$ scattering  as well as in real \gaga\ scattering 
(see discussions at the end); the experimental status of LAC1,2 is less clear. 

The recent \ub{WHIT parametrizations} \cite{41} follow a similar philosophy
as LAC, at least regarding the gluon input; however, their choices for
$G_0^{\gamma}$ are much less extreme. In the WHIT1,2,3 parametrizations, 
gluons carry about half as much of the photon's momentum as quarks do (at the
input scale $Q_0^2 = 4$ GeV$^2$), while in WHIT4,5,6 gluons and quarks carry
about the same momentum fraction and in each set softness of the input 
gluon density was systematically increased. These only exist in LO, but great 
care has been taken to treat the ($x-$dependent) charm threshold correctly. 
This is much more important here than for nucleonic parton densities, since 
the photon very rapidly develops an ``intrinsic charm" component from 
$\gamma \to \ccbar$ splitting.

The \ub{GRV parametrization} \cite{42} is the first NLO fit of \qvph; a LO
version is also available. This parametrization is based on the same
``dynamical" philosophy where one starts from a very simple input 
at a very low $Q_0^2$ (0.25 GeV$^2$ in LO, 0.3 GeV$^2$ in NLO); 
this scale is assumed to be the same for $p, \ \pi$ and $\gamma$ targets. 
The observed, more complex structure is then generated dynamically by the 
evolution equations.  The input densities for the photon are taken 
proportional to those for the (vector meson and hence) pion case\cite{grvpi}.
Over and above the $\gamma \to  \rho$ transistion probability given by the
VDM there is a proportionality factor $\kappa$ which is the {\em only} free
parameter in this ansatz and was determined to $\kappa=2$ (1.6) in LO (NLO).
This approach has met with some criticism due to the low scale used. 
The GRV parametrization ensures a smooth onset of the charm density, 
using an $x-$independent threshold.

The \ub{GS parametrizations} \cite{38} were developed shortly after GRV, but
follow a quite different strategy. Problems with low input scales \cite{38a}
are avoided by choosing $Q_0^2 = 5.3$ GeV$^2$. This is certainly in the
perturbative region, but necessitates a rather complicated ansatz for the
input distributions: 
\be \label{e2.13}
\vec{q}^{\gamma}_{0,{\rm GS}}(x) = \kappa \frac {4 \pi \aem} {f^2_{\rho}}
\vec{q}^{\pi}_0(x,Q_0^2) + \vec{q}^{\gamma}_{\rm QPM}(x,Q_0^2).
\ee
The free parameters in the fit are the momentum fractions carried by gluons
and sea--quarks in the pion, the parameter $\kappa$, and the light quark 
masses. In the GS2 parametrization, $G_0^{\gamma}$ is assumed to come entirely
from the first term in eq.(\ref{e2.13}), while in GS1 the second term also
contributes via radiation. While the fit gives 
reasonable values for all the three parameters, the ansatz (\ref{e2.13}) 
though true in the perturbative region is not invariant under the evolution 
equations.  For practical purposes, however, it includes sufficiently 
many free parameters to allow a decent description of data on 
$F_2^{\gamma}$. The newer version uses of this parametrisation \cite{38b}
uses slightly reduced input scale $Q_0^2=3$ GeV$^2$, and for the first time
includes data on jet production in two--photon collisions  in the
fit; unfortunately this still does not allow to pin down \gga\ with any
precision. 

The \ub{AGF parametrization} \cite{47} is (in its ``standard" form) quite
similar to GRV. In particular, they also assume that at a low input scale
$Q_0^2 = 0.25$ GeV$^2$ the photonic parton densities are described by the 
VDM. The main difference is in the scheme used for determining the input
densities as well as inclusion of the $\rho-\phi-\omega$ interference effects.
Separate fits are provided for the ``anomalous" (or ``pointlike") and 
``non--perturbative" contributions to \qvph, allowing the user to specify the
absolute normalization (although not the shape) of the latter.

Finally, two of the \ub{SaS parametrizations} \cite{36a} are based on a
similar philosophy as the GRV and AGF parametrizations, by assuming that at a
low $Q_0 \simeq 0.6$ GeV the perturbative component vanishes (SaS1). However,
while the normalization of the non--perturbative contribution is taken from
the VDM the shapes of the quark and gluon distributions are fitted
from data. Although the SaS parametrizations are available in LO only, the
authors attempt to estimate the scheme dependence 
providing a parametrization (SaS1M) where the non leading--log part of the QPM
prediction for $F_2^{\gamma}$ has been added to eq.(\ref{e2.7}), while SaS1D
is based on eq.(\ref{e2.7}) alone. There are also two parametrizations
(SaS2D, SaS2M) with $Q_0=2$ GeV; however, in this case the normalization 
of the fitted ``soft" contribution had to be left free. The SaS1 sets 
preferred by the authors are quite similar to AGF; the real significance 
of ref.\cite{36a} is that it carefully describes the properties
of the hadronic state $X$ for both the hadronic and ``anomalous" 
contributions, as needed for a full event characterization. 

\begin{figure}[htb]
\leavevmode 
\begin{center}
\mbox{\epsfxsize=.4\hsize\epsffile{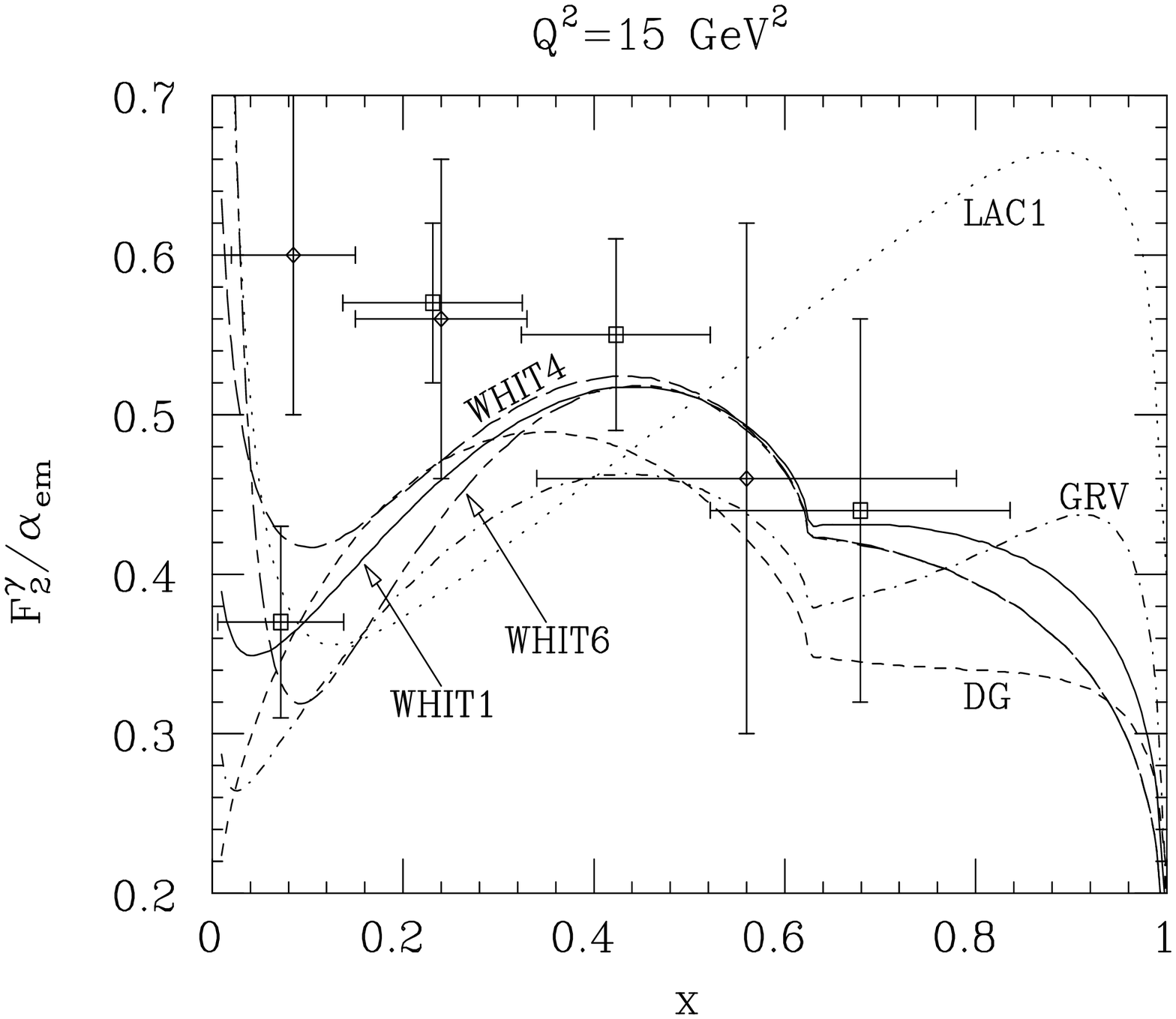}}
\caption{Data on \f2gam\ \protect \cite{28,32} as a fucntion of $x$ 
compared with various parametristions}
\end{center}
\end{figure}
In Fig.~4 we compare various LO parametrizations of $F_2^{\gamma}$ at
$Q^2=15$ GeV$^2$ with recent data taken by the OPAL \cite{28} and TOPAZ
\cite{32} collaborations; present data are not able to distinguish between LO 
and NLO fits. In order to allow for a meaningful comparison, we have added a 
charm contribution to the OPAL data, as estimated from the QPM; this 
contribution had been subtracted in their analysis. We have used the DG 
and GRV parametrizations with $N_f=3$ flavours, since their parametrizations 
of $c^\gamma$ are meant to be used only if $\log Q^2 / m_c^2 \gg 1$; the 
charm contribution has again been estimated from the QPM.
\footnote{We have ignored the small contribution \cite{26} 
from $\gamma^* g \to \ccbar$ in this figure.}
As discussed earlier, WHIT provides a parametrization of \cg\ that includes
the correct kinematical threshold, while LAC treat the charm as massless at
all $Q^2$. 

We see that most parametrizations give quite similar results for 
$F_2^{\gamma}$ over most of the relevant $x-$range; the exception is LAC1,
which exceeds the other parametrizations both at large and at very small $x$.
It should be noted that the data points represent averages over the
respective $x$ bins; the lowest bin starts at $x=0.006 \ (0.02)$ for the OPAL
(TOPAZ) data. The first OPAL point is therefore in conflict \cite{34} with the
LAC1 prediction. Unfortunately there is also some discrepancy between the 
TOPAZ and OPAL data at low $x$. As discussed above, one is sensitive to the
unfolding procedure here; for this reason, WHIT chose not to use these (and
similar) points in their fit. (The other fits predate the data shown in 
Fig.~4.) This ambiguity in present low$-x$ data is to be regretted, since
in principle these data have the potential to discriminate between different
ans\"atze for $G_0^{\gamma}$. This can most clearly be seen by comparing the
curves for WHIT4 (long dashed) and WHIT6 (long--short dashed), which have the
{\em same} valence quark input, and even the same $\int x G_0^{\gamma} dx$:
WHIT4 has a harder gluon input distribution, and therefore predicts a larger
$F_2^{\gamma}$ at $x \simeq 0.1$; WHIT6 has many more soft gluons, and
therefore a very rapid increase of $F_2^{\gamma}$ for $x \leq 0.05$, not 
unlike LAC1. Finally, it should be mentioned that the GS, AGF and SaS
parametrizations also reproduce these data quite well.

\begin{figure}[htb]
\leavevmode 
\begin{center}
\mbox{\epsfxsize=0.4\hsize\epsffile{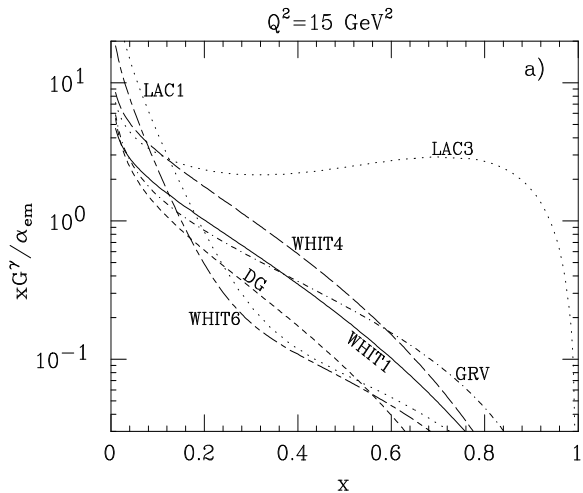}}
\caption{Gluon densities in various parametrisations of \qvph\ }
\end{center}
\end{figure}
Discriminating between these parametrizations would be much easier if one
could measure the gluon density directly. This is demonstrated in Fig.~5,
where we show results for $x \Gg$ at the same value of $Q^2$; we have
chosen the same LO parametrizations as in Fig.~4, and included the LAC3
parametrization with its extremely hard gluon density. Note that, for
example, WHIT4 and WHIT6 now differ by a factor of 5 for $x$ around 0.3.
The gluon distribution of WHIT6 is rather similar in shape to the one of LAC1,
but significantly smaller in magnitude. Indeed, in all three LAC 
parametrizations, gluons carry significantly more momentum than quarks for 
$Q^2 \leq 20$ GeV$^2$; this is counter--intuitive \cite{38}, since in known
hadrons, and hence presumably in a VMD--like low$-Q^2$ photon, gluons and
quarks carry about equal momentum fractions, while at very high $Q^2$ the
inhomogeneous evolution equations (\ref{e2.8}) predict that quarks in the
photon carry about three times more momentum than gluons. Notice finally that
GRV predicts a relatively flat gluon distribution. This results partly 
from the
low value of the input scale $Q_0^2=0.25$ GeV$^2$, compared to 1 GeV$^2$ for
DG and 4 GeV$^2$ for WHIT and LAC1; a larger $Q^2/Q_0^2$ allows for more
radiation of relatively hard gluons off large$-x$ quarks. 

Since the measurement of \f2gam\ can not constrain the flavour structure,
the different parametrisations mentioned above difer substantially from 
each other in their flavour structure as well.

\vspace{0.5cm}

\noindent{\bf \large Hard processes  other than the DIS and \qvph\ : }

\vspace{0.5cm}

The discussion at the end of the last section clearly shows that while the
data on \f2gam\ accumulated till now indeed supports the theoretical
predictions, the DIS data cannot discriminate between the various 
parametristions which differ considerably in their gluon content and 
flavour structure.  In order to make use of the structure function
language effectively to caclculate processes involving photons, we need to 
inprove upon this knowledge. Hard processes where the partons in the
photon participate in the hard scattering, the so called `resolved processes' 
\cite {5}  hold the promise of  being able to do that and an experimental
study of these processes has taken the centre stage in \gaga\ and \gamp\ 
physics  in the last 3-4 years. Both the \gaga\
scattering at TRISTAN and LEP as well as \gamp\ scattering at HERA, has 
begun to provide a lot of data on jet production as well as heavy 
flavour production which have demonstrated ability to discriminate
between different parametrisations of \glph\ . See ref. \cite{5} 
for a summary of the recent developements in the area. Here, 
\begin{figure}[htb]
\leavevmode 
\begin{center}
\mbox{\epsfxsize=0.45\hsize\epsffile{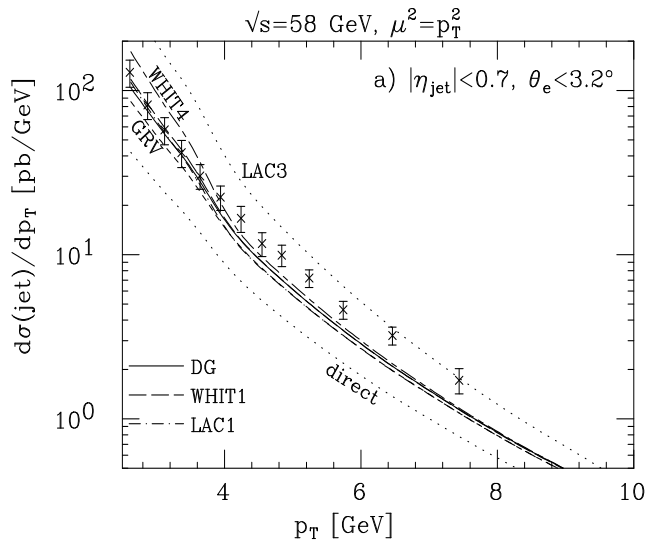}}
\caption{Single--jet inclusive cross--section in \gaga\ collisions obtained 
by TOPAZ \protect\cite{topazjt} compared  with predictions for various 
parametrisations of \qvph }
\end{center}
\end{figure}
I present only as an example of the available experimental information
the inclusive single  jet spectrum expected for the the anti--tagging
conditions of the TOPAZ detector at TRISTAN (the TRISTAN data on 
jet production are  the only published  data where the detector effects 
have been unfolded) for various parametrisations mentioned above. 
The lower dotted curve shows the direct contribution only. Thus the data
clearly demonstrate existence of the `resolved' contributions. The other   
curves show LO predictions for the various parametrisations. It can be 
clearly seen that the data have some discriminatory power and rule out
already the LAC3 parametrisation. The data from TRISTAN on heavy flavour
(charm) production seem to disfavour DG parametrisation  somewhat,
whereas the jet production data from HERA also rule out LAC3. Further
studies of jet and heavy flavour production at HERA, LEP as well as direct
photon production at HERA can indeed provide some more information
on the photonic parton densities.

\vspace{0.5cm}

\noindent{\bf \large Conclusions:}

\vspace{0.5cm}
\begin{itemize}
\item[1]  The basic predictions of perturbative QCD as regards the \qsq\
and $x$ dependence of \f2gam\ have been confirmed by experiments.  The
only sensible way is to treat the \f2gam\ similar to the nucleon structure
function  and fit  the form of input densities at a low scale, using the 
data  on \f2gam\ and the evolution equations. 
\item[2] Various parametrisations of the photonic parton densities \qvph\
exist all of which describe the data on \f2gam\ well, but they differ a lot 
in the gluon densities \glph\ as well as in their  flavour strucutre.
Data at small $x$ from  LEP2 might be able to distinguish between 
different parametrisations.
\item[3] The `Resolved Photon Processes', where the partons in the photon
participate in the hard scattering  also has the potential of providing
important information about \qvph\ in general and the gluon density in 
particular. Data from HERA ($ep$ experiments) and TRISTAN/LEP ($e^+e^-$
experiments) have already confrimed the existence of the `resolved photon' 
processes at the expected level and have begun to provide nontrivial 
information on \qvph\ . 

\end{itemize}

\end{document}